\documentclass[a4paper,12pt]{article}
\usepackage{amssymb}
\usepackage{eucal}
\usepackage{pstricks,pst-node,pst-plot}
\usepackage{graphicx}
\newtheorem{theorem}{Theorem}[section]

\newtheorem{cor}[theorem]{Corollary}
\newtheorem{prop}[theorem]{Proposition}

\def\Sch{Schr\"odinger }

\def\eref#1{(\ref{#1})}
\def\dd{\partial}
\def\half{\frac{1}{2}}

\def\be{\begin{equation}}
\def\eeq{\end{equation}}
\def\bea{\begin{eqnarray}}

\def\eea{\end{eqnarray}}
\def\ba{\begin{array}{l}}
\def\ea{\end{array}}
\def\eel#1{\label{#1}\end{equation}}
\def\eeal#1{\label{#1}\end{eqnarray}}
\def\a{\alpha}
\def\l{\lambda}
\def\d{{\rm d}}
\def\dd{\partial}
\def\pdd#1#2{\frac{\dd #1}{\dd #2}}

\def\CM{Calogero-Moser }
\def\BT{B\"acklund transformation }

\def\DT{Darboux transformation }
\def\DTs{Darboux transformations }
\def\Schr{Schr\"odinger }

\def\efn{eigenfunction }

\def\evals{eigenvalues }
\def\qpoly{quasi-polynomial }
\def\qrat{quasi-rational }

\title{On the rational monodromy-free potentials with sextic growth}

\author{J.Gibbons\\Imperial College, 180 Queen's Gate,\\ London SW7 2BZ\\j.gibbons@ic.ac.uk
\and\\ A.P. Veselov\\Department of Mathematics, Loughborough University,\\
Loughborough, LE11 3TU \\and
\\Landau Institute for Theoretical Physics,\\
Moscow, Russia\\A.P.Veselov@lboro.ac.uk}
\begin{document}
\maketitle
\begin{abstract}

We study the rational potentials $V(x)$, 
with sextic growth at infinity, such
that the corresponding one-dimensional \Sch equation has no monodromy in
the complex domain for all values of the spectral parameter. We investigate
in detail the subclass of such potentials which can be constructed by the
Darboux transformations from the well-known class of quasi-exactly 
solvable potentials
$$V= x^6 - \nu x^2 +\frac{l(l+1)}{x^2}.$$ 
We show that, in contrast with the case of quadratic growth, there are 
monodromy-free potentials which have quasi-rational eigenfunctions, 
but which can not be given by this construction. 
We discuss the relations 
between the corresponding algebraic varieties, and present some 
elementary solutions of the Calogero-Moser problem in the external 
field with sextic potential.

\end{abstract}

\section{Introduction}

Consider the Schr\"odinger equation with a meromorphic
potential $V(x)$:
\begin{equation}
\label{static}
(-D^2+V(x)) \psi = \lambda \psi,
\end{equation}
where here and below $D= \frac{\mathrm d}{{\mathrm d}x}$.
We
say that such a potential $V(x)$ has {\it trivial monodromy}, 
if all the 
solutions $\psi$ of the corresponding equation (\ref{static}) are 
meromorphic in $x$ for {\it all} $\lambda$.
The general problem is to describe all such potentials.

In such generality this is probably a hopeless question. 
However if we restrict ourselves to a suitable class of potentials it 
is known to be solvable.
The first results in this direction were found by
Duistermaat and Gr\"unbaum  \cite{DG}, who solved it in the class of 
rational potentials decaying at infinity:
\[
V=\sum_{i=1}^N\frac{k_i(k_i+1)}{(x-x_i)^2}.
\]
They showed that the corresponding parameters $k_i$ must be integers, 
and all such potentials are the results of Darboux transformations 
applied to the zero potential. Therefore they are given by the Burchnall-Chaundy (or Adler-Moser) explicit formulas \cite{BCh, AdMos}. 
The corresponding configurations of the poles $x_i$ are very special: in the case when all the parameters $k_i=1$, they are nothing else but the (complex) equilibria of the Calogero-Moser system, with the Hamiltonian
\[
H=\sum_{i=1}^N p_i^2 +
\sum_{j \neq i}^N\frac{2}{(x_i-x_j)^2},
\]
which are described by the following algebraic system:
\begin{equation}
\label{eCM}
 \sum_{j\ne i}^N\frac{ 1}{(x_i - x_j)^3}=0, \quad i=1,\dots, N.
\end{equation}
A remarkable fact (discovered first by Airault, McKean and Moser \cite{AMM}) is that this system has no solutions unless 
\[
N = \frac{m(m+1)}{2}
\] 
is a triangular number, in which case the solutions depend on $m$ arbitrary complex parameters. 

Oblomkov \cite{Oblomkov} generalised the Duistermaat-Gr\"unbaum result to the case of rational potentials with quadratic growth at infinity:
\[ 
V=x^2 + \sum_{i=1}^N\frac{k_i(k_i+1)}{(x-x_i)^2},
\] 
by proving that all such potentials can be found by applying successive 
Darboux transformations to the harmonic oscillator, and that they have 
the following explicit description in terms of the Wronskians of 
Hermite polynomials $H_k(x)$:
\[
V=x^2 - 2D^2 \log W(H_{k_1}, H_{k_2},\dots, H_{k_n}).
\]
In particular the corresponding Schr\"odinger equations must have 
infinitely many solutions of the  {\em quasi-rational} form
\begin{equation}
\label{qrs}
\psi = R(x) \exp(\pm x^2/2),
\end{equation}
where $R(x)$ is some rational function. 

An interpretation of this result in terms of the generalised {\it Stieltjes relations} and the rational solutions of the dressing chain \cite{VS} was proposed in the paper \cite{VesSti}, where also
a new class of meromorphic monodromy-free potentials, expressible in terms of the Painleve transcendents, was found.

In this paper we consider the set $\mathbf M$ of monodromy-free 
even rational potentials with {\em sextic} growth at infinity:
\begin{equation} 
\label{sextic}
V=P_6(x) + \sum_{i=1}^N\frac{k_i(k_i+1)}{(x-x_i)^2}
\end{equation} 
where $P_6(x)$ is an even polynomial of degree 6, which for simplicity 
will be assumed to be of the form
\begin{equation}
P_6(x) = x^6 - \nu x^2,
\end{equation} 
where $\nu$ is a parameter. 

Our motivation came from the theory of Calogero-Moser systems. It is known,
after Inozemtsev \cite{Inoz}, that the corresponding classical system in
an external field with the Hamiltonian:
\[
H=\sum_{i=1}^N p_i^2 + \sum_{i=0}^N P_6(x_i) + 
\sum_{j \neq i}^N\frac{1}{(x_i-x_j)^2}
\]
is integrable (note that this is believed to be not the case if the degree
of $P$ is higher than 6), so one can expect some effective way to describe
the equilibria of this system and the corresponding set $\mathbf M.$

Having in mind the case of quadratic growth, 
let us consider also the set 
$\mathbf S$ of potentials (\ref{sextic}), such that the corresponding
Schr\"odinger operator has a {\em quasi-rational} eigenfunction, of the form
\begin{equation}
\label{qrf}
\psi = R(x) \exp(\pm x^4/4),
\end{equation}
where $R(x)$ is some rational function. 

Finally we consider the set $\mathbf D$ of the operators, which
are Darboux-related to the operators of the form
\begin{equation}
\label{qrs}
\mathcal L_{p,m}= -D^2 + x^6 - (2p+1)x^2 + \frac{m(m+1)}{x^2}
\end{equation}
with integer $p$ and $m.$ The operators $\mathcal L_{p,m}$ are known, after
Turbiner, \cite{Turb1, Turb2}, to be {\it quasi-exactly solvable}, and they
were studied later from various points of view in \cite{ChhajMal, ChhajLetMal,
TatTurb, BD, KR, DDT}. They have $M$ quasi-rational eigenfunctions
\[
\psi_i ^{(p,m)}= R_i^{(p,m)}(x) \exp(\pm x^4/4), \quad i=1,\dots, M, 
\]
where $M$ is a positive integer determined by $p$ and $m$ (see section 2
below). The set $\mathbf D$
consists of the potentials of the form:
\begin{equation}
\label{qrl}
V(x) = x^6 - (2p+1)x^2 + \frac{m(m+1)}{x^2} -2 D^2 \log W_I(x),
\end{equation}
where $I \subset \{1,2, \dots, M\}$ and $W_I$ is the Wronskian of the 
corresponding quasi-rational eigenfunctions $\psi_i, \, i \in I.$ 

We will study all these three sets and the corresponding affine algebraic
varieties. In particular, we show that, under some assumptions, 
$$\mathbf D \subset \mathbf S \subset \mathbf M,$$ 
and that, in contrast to the quadratic case, all the inclusions are 
proper: 
$$\mathbf D \neq \mathbf S \neq \mathbf M,$$
even if we restrict the definition of the set $\mathbf M$ to odd integer
values of the parameter $\nu$. This means that only a proper subset of 
monodromy-free potentials with sextic growth at infinity have an 
explicit description in terms of Darboux transformations.

We study also the corresponding time-dependent \Sch equation, and show that
the Stieltjes relations for them can be considered as B\"acklund transformations
between Calogero-Moser systems in different sextic potentials, with different
numbers of particles. We construct a family of explicit solutions
in elementary functions for such a system. We note that these solutions are
very special - Inozemtsev's general solution of the system is highly 
transcendental, being expressed in terms of 
Riemann $\theta$-functions.


\section{The locus conditions}

In this section we recall the trivial monodromy conditions \cite{DG}, 
which we will call {\it locus equations} following the terminology of \cite{AMM}.

Suppose that a \Schr equation with a rational
potential $V(x)$
\[
( - D^2+V(x)) \psi = \lambda \psi,
\]
has all its solutions meromorphic for all $\lambda$.
At a singular point, which we may take to be  $x=0$, we have the two
series: 
\begin{eqnarray*}
V(x)=x^{-D}\sum_{n=0}^\infty V_n x^n,\\
\psi(x,\lambda)= x^{-m} \sum_{n=0}^\infty \psi_n x^n.
\end{eqnarray*}
where we suppose $D,m>0$, and $\psi_0\ne 0$. Then the \Schr equation becomes:
\begin{eqnarray*}
( -D^2
+\sum_{n=0}^\infty V_n x^n) \sum_{l=0}^\infty \psi_l x^{l-m}\\ 
= \lambda x^{-m} \sum_{n=0}^\infty \psi_n x^n.
\end{eqnarray*}
Expanding, we find:
\begin{eqnarray*}
-\frac{1}{2} \sum_{n=-(m+2)}^\infty (n+2)(n+1) \psi_{n+2} x^n + x^{-(m+D)}
\sum_{n=0}^{\infty}x^n
\sum_{l=0}^{n} \psi_l V_{n-l}\\
=\lambda x^{-m} \sum_{n=0}^\infty \psi_n x^n.
\end{eqnarray*}
Hence, to balance the leading terms, it follows that $V(x)$ has only double
poles, $D=2$, and that $V_{-2} = \frac{1}{2} m(m+1)$. The integer $m$ will
be called the {\em multiplicity} of the pole.

There are further constraints. The coefficient of $x^n$ includes the 
term 
$$\frac{1}{2}(m(m+1)-n(n-1))\psi_n$$
and other terms which involve $\psi_l V_{n-l}$, for $l<n$. Hence the $m+1$st
equation does not involve $\psi^{m+1}$. Thus the \Schr equation
can be solved to all orders, provided that the $(m+1)$st equation 
is consistent with the previous equations. 
The resulting system may be written in matrix form:
\[
\left(\matrix{ 0& \ldots&\ldots&\ldots & m&V_1\cr
               0& \ldots&\ldots&2m-1&V_1&V_2-\lambda\cr
               0&\ldots&\ldots&\ldots&\ldots&\ldots\cr
               0&\ldots&ml-l(l-1)/2&V_1&\ldots&V_l\cr
               0&\ldots&\ldots&\ldots&\ldots&\ldots\cr
               0& 2m-1&V_1&V_2-\lambda&\ldots&V_{2m-1}\cr
               m&V_1&V_2-\lambda&\ldots&&V_{2m}\cr
               V_1&V_2-\lambda&\ldots&&V_{2m}&V_{2m+1}}\right) 
               \left(\matrix{ \psi_{2m}\cr\psi_{2m-1}\cr\cdot\cr\cdot\cr\psi_1\cr\psi_0}\right)
               =\left(\matrix{0\cr0\cr\cdot\cr\cdot\cr0\cr0}\right).    
\]
The $(2m+1)\times(2m+1)$ matrix on the left must be singular for all 
values of $\lambda$; it follows from this that the leading $m+1$ odd 
coefficients of the Laurent series of $V$ all vanish:
\be
\label{locus}
V_1=V_3=\ldots=V_{2m+1}=0.
\eeq

In the special case of potentials of the form 
\[ 
V=P(x) + \sum_{i=1}^N\frac{k_i(k_i+1)}{(x-x_i)^2},
\]
where $P(x)$ is some polynomial, we have an algebraic system of 
$K= \sum_{j=1}^N k_j$ equations in $N$ variables $x_i$:
\begin{equation}
P^{(2s-1)}(x_i) 
- (2s)!  \sum_{j\ne i}^{N}\frac{k_j(k_j+1)}{(x_i - x_j)^{2s+1}} =0,
 \quad i=1,\dots, N,\, s=1,\dots k_i.
\end{equation}
If all the multiplicities $k_i=1$ we have $N$ equations for $N$ variables:
\begin{equation}
\label{locusgen}
P'(x_i) -  \sum_{j\ne i}^{N}\frac{2}{(x_i - x_j)^{3}} =0, 
\quad i=1,\dots, N.
\end{equation}

In our case, with $P(x) = x^6 - \nu x^2$, we will consider another 
special case, when $x=0$ may have an arbitrary  multiplicity $l$, 
all other poles have multiplicity $1$, and the potential is even:
\begin{equation}
\label{gen}
V(x) = x^6 - \nu x^2 + \frac{l(l+1)}{x^2} + \sum_{i=1}^{N}\frac{2}{(x-x_i)^2},
\end{equation}
where 
$x_1, \dots, x_{N},\, N=2n$ is a set of distinct non-zero complex numbers
symmetric with respect to the origin. In that case the locus equations at
$x=0$ are automatically satisfied, and we have the following algebraic system
on $x_i$:
\begin{equation}
\label{locus}
 \sum_{j\ne i}^{N}\frac{2}{(x_i - x_j)^3} + \frac{l(l+1)}{x_i^3} 
 + \nu x_i - 3x_i^5 =0, \quad i=1,\dots, N.
\end{equation}
This is an affine algebraic variety $\mathbf M^{(N)}(\nu,l)$, 
which is non-empty for all values of $N$ and the parameter $\nu$.  

What we are going to show here is that, for special values of the parameter
$\nu$, there is a subset of such equilibria which can be described quite
explicitly. They correspond to a special subclass of the monodromy-free 
potentials  (\ref{sextic}),
and form some zero-dimensional algebraic varieties, whose properties we would
like to study.

\section{The Darboux transformation}

We recall here the classical procedure going back to Darboux \cite{Darb}.

Let us consider the Schrodinger operator $L^{(0)}$ with potential 
$V^{(0)}(x)$, its eigenfunctions $\psi^{(0)}$, 
and in particular its eigenfunction
$\psi^{(0)}_1$ with eigenvalue $\lambda_1$:
\be
(L^{(0)}-\lambda_1)\psi^{(0)}=( -D^2+V^{(0)}(x)-\lambda_1) \psi^{(0)} 
= (\lambda-\lambda_1)\psi^{(0)}.
\label{sch0}
\eeq
The \Sch operator can be factorised as
\[
( -D^2+V^{(0)}(x)-\lambda_1) 
=  -(D+v^{(0)})(D-v^{(0)}).
\]
The operator on the right must annihilate the \efn $\psi^{(0)}_1$,  
so we may take $v^{(0)} = D\ln(\psi^{(0)}_1)$. The
potential $V^{(0)}(x)$ is given by
\[
V^{(0)}(x)-\lambda_1 = \frac{\d v^{(0)}}{\d x}+{v^{(0)}}^2.
\]
Now we act on the left of \eref{sch0} with the operator $(D-v^{(0)})$. 
This takes
any eigenfunction $\psi^{(0)}$ of $L^{(0)}$ into a new function 
$$\psi^{(1)}=(D-v^{(0)})\psi^{(0)}$$
which is either zero, if $\psi^{(0)}$ is a multiple of $\psi^{(0)}_1$,
or else an eigenfunction of a new operator $L^{(1)}$. Explicitly
\[
-(D-v^{(0)})(D+v^{(0)})(D-v^{(0)})\psi^{(0)}
= (\lambda-\lambda_0)(D-v^{(0)})\psi^{(0)}.
\]
so that
\be
(L^{(1)}-\lambda_1)\psi^{(1)}
=-(D-v^{(0)})(D+v^{(0)})\psi^{(1)}
= (\lambda-\lambda_1)\psi^{(1)}.
\eeq
The potential $V^{(1)}(x)$ in the transformed operator 
\[
L^{(1)}-\lambda_1=-(D-v^{(0)})(D+v^{(0)})
\]
is then given by
\[
V^{(1)}(x)-\lambda_1 =- \frac{\d v^{(0)}}{\d x}+{v^{(0)}}^2.
\]
Thus
\be
V^{(1)}(x)-V^{(0)}(x) = -\frac{\d v^{(0)}}{\d x}=-\frac{\d^2}{\d x^2}\ln{\psi^{(0)}_1}.
\eeq

If we now choose one of the remaining eigenvalues $\l_2$, say, 
then the transformed Schr\"odinger equation, 
shifted by this eigenvalue, is
\be
(L^{(1)}-\lambda_1)\psi^{(1)}=( -D^2+V^{(1)}(x)-\lambda_1) \psi^{(1)} 
= (\lambda-\lambda_1)\psi^{(1)}.
\label{sch1}
\eeq
We may now repeat this procedure indefinitely. The corresponding operators
$L^{(n)}$ are related by
\[
L^{(n)}D^{(n)} = D^{(n)}L^{(0)},
\]
where the $n$-th order `dressing operator' is given by:
\[
D^{(n)}=(D-v^{(n-1)})(D-v^{(n-2)})\ldots(D-v^{(0)}).
\]
Since $D^{(n)}$ must annihilate the chosen set of eigenfunctions 
\[
\left\{\psi^{(0)}_1, \ldots, \psi^{(0)}_n\right\}
\] 
of $L^{(0)}$, with respective \evals $ \l_1,\l_2 \ldots, \l_n$, we have
\[
\psi^{(n)}=D^{(n)}\psi^{(0)} =\frac{ W_{n+1}(\psi^{(0)}_1, \psi^{(0)}_2,
\ldots, \psi^{(0)}_n,\psi^{(0)})}
{ W_{n}(\psi^{(0)}_1, \psi^{(0)}_2, \ldots, \psi^{(0)}_n)},
\]
where $W_n$ denotes the Wronskian of its $n$ arguments. Further, it may
be shown by induction (see Crum \cite{Crum}) that
\bea
V^{(n)}(x) = 
V^{(0)}(x)-2 \frac{\d^2}{\d x^2}\ln{W_{n}(\psi^{(0)}_1, \psi^{(0)}_2, 
\ldots, \psi^{(0)}_n}).
\label{Crum}
\eea

What is important for us, is that if the initial potential 
 $V^{(0)}$ is meromorphic and has trivial monodromy, then it follows from
 (\ref{Crum}) that the same is true
for $V^{(n)}$ as well. We will use this fact in the next section
to describe a class of monodromy-free potentials with sextic growth at infinity.

\section{The even sextic potential}

The anharmonic oscillator in an even sextic potential, with
Hamiltonian
\[
H=\frac{1}{2} (p^2 + x^6 - \nu x^2)
\]
has the special property that the corresponding system
\[
\frac{d^2 x}{dt^2} = -3 x^5 + \nu x
\] 
can be solved explicitly using the elliptic Weierstrass $\wp$-functions:
\[
x(t)=\frac{1}{\sqrt{\wp(\sqrt{2E}(t-t_0);g_2,g_3)-\frac{\nu}{6\,E}}},
\]
where the invariants $g_2$ and $g_3$ are given by:
\begin{eqnarray*}
g_2=\frac{\nu^2}{3 E^2}\\
g_3=\frac{2}{E}-\frac{\nu^3}{27 E^3},
\end{eqnarray*}
and the energy $E$ and starting time $t_0$ are the two arbitrary constants.

In contrast, the Schr\"odinger equation for the 
corresponding quantum system
\be
\left(-D^2 + x^6 - \nu x^2\right)\psi 
= \lambda \psi \label{Sch6}
\eeq
can not in general be solved exactly; 
instead it is at best {\it quasi-exactly solvable}
in the sense of Turbiner \cite{Turb2}. 

To be more precise, let us consider the following slightly more general 
family of Schr\"odinger equations
\cite{Turb2}
\begin{equation} 
(-D^2 + x^6 - \nu x^2 +\frac{l(l+1)}{x^2})\psi 
= \lambda \psi, \label{Sch7}
\end{equation}
studied in detail in particular in 
\cite{ChhajMal, ChhajLetMal, TatTurb, BD, KR, DDT}.
Let us look for a special class of {\it quasi-polynomial solutions} 
of the form:
\be
\psi = x^{\mu}P(x) \exp(\pm x^4/4), \label{qps}
\eeq
where $P(x)$ is a polynomial.
Let us consider first the case when
\be
\psi = x^{\mu}P(x) \exp(- x^4/4). \label{qps-}
\eeq
It is useful to rewrite this equation in terms of $\phi,$ where 
$\psi = \phi(x) \exp(-x^4/4)$:
\[
 -( D^2 -2 x^3  D
        - (\nu - 3) x^2  + \frac{l(l+1)}{x^2}) \phi = \lambda \phi.
\]
The coefficients $a_n$ of a series expansion for $\phi$:
\[
\phi = x^{\mu}\sum_{n=0}^\infty a_n x^n,
\]
satisfy the following recurrence relation
\be
\label{recurr}
 [(\mu+n+2)(\mu+n+1)-l(l+1)] a_{n+2} +  \lambda a_n +(\nu-2n-2\mu+1) a_{n-2} =0.
\eeq
The very first equation gives
\[
\mu(\mu-1)- l(l+1)=0,
\]
which implies that either $\mu=l+1$ or $\mu=-l$.
In the first case, the recurrence becomes:
\be
\label{firs}
(n+2)(n+2l+2) a_{n+2} +  \lambda a_n +(\nu-2n-2l-1) a_{n-2} =0,
\eeq
in the second:
\be
\label{sec}
(n+2)(n-2l+1) a_{n+2}  +  \lambda a_n +(\nu-2n+2l+1) a_{n-2} =0,
\eeq
In particular, we see 
that the even and odd-index terms are decoupled, which is a consequence
of the symmetry of the original problem. The quasi-polynomial solutions
exist when $\nu-2n-2l-1=0$  for some even positive integer $n$: 
in that case the $a_n$ can all be set to zero for $n \geq 2M,$
where
$$
M=\frac{\nu-(2l+1)}{4},
$$
which in that case is a positive integer.
The eigenvalues $\l$ must satisfy the characteristic
equation of the corresponding tridiagonal matrix, which is an algebraic 
equation (with integer coefficients if $l$ is an integer) of degree 
$2M-2.$ These polynomials are studied in more detail by Bender and 
Dunne \cite{BD} and sometimes called {\it Bender-Dunne polynomials}.

When $\mu=-l$ we have $M$ quasi-polynomial solutions (\ref{qps-}), 
where 
$$
M=\frac{\nu+(2l+1)}{4}
$$
provided $M$ is a positive integer.

Similarly, the quasi-polynomial solutions of the form
\be
\psi = x^{l+1}P(x) \exp( x^4/4), \label{qps+}
\eeq
exist when
$$
M=\frac{-\nu-(2l+1)}{4}
$$
is a positive integer; in that case there are $M$ such solutions.
For the quasi-polynomal solutions of the form
\be
\psi = x^{-l}P(x) \exp( x^4/4), \label{qps+}
\eeq
the same is true for 
$$
M=\frac{-\nu+(2l+1)}{4}.
$$

Summarising the above, we have: 
\begin{prop}
The Schr\"odinger equation (\ref{Sch7}) with integer $l$ 
may have quasi-polynomial solutions of the form (\ref{qps})
only if $\nu$ is an odd integer. In that case the number 
of such solutions is $M$, which is the non-negative integer 
of the form
\be
M=\frac{\pm\nu\pm(2l+1)}{4}.\label {M}
\eeq 
The combination of the signs in this formula for $M$ determines the 
type of the corresponding solutions
(\ref{qps}), where $P(x)$ must be an even polynomial of degree $2M-2$. 
The corresponding eigenvalues $\lambda_1, \dots, \lambda_M$ are 
distinct real algebraic numbers. 
\end{prop}

In particular, we see that such solutions exist for all odd $\nu$ except
$\nu=\pm(2l+1).$

{\bf Example.} Let $l=0,\, \nu = 7$:
\be
L= -D^2 +  x^6 - 7 x^2.
\eeq
Then $M=\frac{7+1}{4}=2.$ Both signs are plus, so the corresponding 
quasi-polynomial solutions are of the form 
$$
\psi = x^{-l}P(x) \exp(- x^4/4) = P(x) \exp(- x^4/4), \, P(x)=a_0 + a_2 x^2.
$$
The characteristic equation  is
\be
\left|\matrix{ 
  \lambda & 2 \cr 
 4 &  \lambda}\right| =0,
\eeq
giving $\lambda = \pm 2 \sqrt{2}$. For  $\lambda = - 2\sqrt{2}$, the eigenfunction
is found to be
\be
\psi_1 = (\sqrt{2} x^2+1) \exp(-x^4/4)
\eeq
which, having no zeroes, must be the ground state, 
while for $\lambda = 2\sqrt{2}$, the eigenfunction
is
\be
\psi_2 = (\sqrt{2} x^2-1) \exp(-x^4/4).
\eeq
This has two zeroes, so there must be another (odd) excited state, with a
single zero, with eigenvalue
between in the interval $( -2 \sqrt{2},2\sqrt{2})$ - however this eigenfunction
is not quasi-polynomial.


\section{Dressing the sextic potential}

The potentials
$$V= x^6 - \nu x^2 +\frac{l(l+1)}{x^2}$$
with integer $l$ are clearly monodromy-free, so we can apply the Darboux
transformations to construct more operators with trivial monodromy. If we
use only quasi-polynomial eigenfunctions, all the new potentials will 
be rational with sextic growth at infinity. 
Indeed by the Crum formula (\ref{Crum}) the new potential is
$$V_I = V -2 D^2 \ln W_{I}(x),$$
where $W_I = W(\psi_{i_1}, \psi_{i_2}, \ldots, \psi_{i_m})$ is the 
Wronskian of the quasi-polynomial eigenfunctions (\ref{qps}) from a 
subset $I=\{i_1, i_2, \dots, i_m\} \subset \{1, \dots, M\}$.

In the case when these eigenfunctions have the form
\be
\psi_i(x) = x^{l+1} P_i(x) \exp( x^4/4),
\eeq
we have the following 

\begin{prop}
The Wronskian $W_I$ has a form
\be
\label{Wron}
W_I = x^{m(l+1) + \half m(m-1)} P_{I}(x) \exp( m x^4/4)
\eeq
with an even polynomial $P_{I}$ of degree 
\be
2d=2m(M-m)
\eeq
 non-vanishing at zero.
 \end{prop}
 
 Indeed we have 
\[W(\psi_{i_1}, \psi_{i_2}, \ldots, \psi_{i_m})=
x^{m(l+1)} \exp( m x^4/4) W(P_{i_1}, P_{i_2}, \ldots, P_{i_m}),
\]
where $P_i$ are the corresponding even polynomials whose coefficients $a_k$
are determined recurrently by (\ref{recurr}). We claim that
\be
\label{mm}
W(P_{i_1}, P_{i_2}, \ldots, P_{i_m}) = x^{\half m(m-1)} P(x)
\eeq
for some even polynomial $P$ with $P(0) \neq 0.$
To prove this let us assume without loss of generality that $i_k=k.$ 
We can apply elementary row operations to the coefficients of the 
polynomials $P_1, \dots, P_m$
so that the lowest order terms are of increasing degree:
$$\tilde P_k = x^{2k} Q_k(x)$$ for some even polynomials $Q_k(x)$. 
We claim that $Q_k(0) \neq 0.$
For this we need to prove the independence of the corresponding vectors 
\[
a^{(i)}= (a^{(i)}_0, a^{(i)}_2, \dots, a^{(i)}_{2k}), \, i=1,\dots, k
\] 
determined by the recurrence relations (\ref{firs}). 
From these relations it follows that
$a_{2j}$ is a polynomial in $\lambda$ of degree $j$ and thus 
$a^{(i)}_{2j}$ are the values of these polynomials at the 
eigenvalues $\lambda_i$. 
Since we know that all $\lambda_i$ are distinct the claim now follows from
the Vandermonde formula.

This implies that the leading term of the Wronskian 
\[
W(P_{i_1}, \ldots, P_{i_m}) = W(\tilde P_{i_1}, \ldots, \tilde P_{i_m}) 
\]
at zero is 
\[x^{0+1+\dots +(m-1)} = x^{\half m(m-1)}.\]

Applying a similar procedure at infinity we can instead make the degrees strictly {\em decreasing}:
$$deg \tilde P_i = 2M-2i,$$ so the degree of the Wronskian is 
$$
2M-2 +(2M-5)+(2M-8)+\ldots+(2M-3m+1)= m(2M-2) -\frac{3}{2}m(m-1).
$$
This means that the degree of the polynomial $P(x)$ in (\ref{mm}) is
$$m(2M-2) -\frac{3}{2}m(m-1)- \frac{1}{2}m(m-1) = 2m(M-m).$$
This completes the proof.

Thus the result of the corresponding Darboux transformations has the form
$$
V_I=x^6 - (\nu-6m)x^2 + \frac{l(l+1) + 2m(l+1) + m(m-1)}{x^2} - 2 D^2 \ln P_I(x) ,
$$
which is equal to
\be
\label{D+-}
V_I=x^6 - (\nu-6m)x^2 + \frac{(l+m)(l+m+1)}{x^2} - 2 D^2 \ln P_I(x)
\eeq
with an even polynomial $P_I$ of degree $2d=2m(M-m)$, where
$$M = \frac{\nu - (2l+1)}{4}.$$
We will denote this set of potentials $D_{+-}$ having in mind the combination of signs in the formula for $M.$

Similarly we have three more sets:

$D_{++}$ consisting of the results of $m$ Darboux transformations using the eigenfunctions of the form
$\psi_i(x) = x^{-l} P(x) \exp{-x^4/4}:$
$$
V_I= x^6 - (\nu-6m)x^2 + \frac{(l-m)(l-m+1)}{x^2} - 2 D^2 \ln P_I(x),
$$
$$deg P_I(x) = 2m(M-m),\, M = \frac{\nu + (2l+1)}{4};$$

$D_{-+}$ consisting of the results of $m$ Darboux transformations using the eigenfunctions of the form
$\psi_i(x) = x^{-l} P(x) \exp{x^4/4}:$
$$
V_I= x^6 - (\nu+6m)x^2 + \frac{(l-m)(l-m+1)}{x^2} - 2 D^2 \ln P_I(x),
$$
$$deg P_I(x) = 2m(M-m),\, M = \frac{-\nu + (2l+1)}{4};$$

$D_{--}$ consisting of the results of $m$ Darboux transformations using the eigenfunctions of the form
$\psi_i(x) = x^{l+1} P(x) \exp{x^4/4}:$
$$
V_I= x^6 - (\nu+6m)x^2 + \frac{(l+m)(l+m+1)}{x^2} - 2 D^2 \ln P_I(x),
$$
$$deg P_I(x) = 2m(M-m),\, M = \frac{-\nu - (2l+1)}{4}.$$

However there are two natural {\bf duality relations} between these sets. They correspond to the spectral equivalences between the operators (\ref{Sch7}) studied by Dorey, Dunning and Tateo in \cite{DDT}.

First of all the form of the potentials in (\ref{Sch7}) is clearly invariant under the change
\be
\label{dual1}
l \rightarrow -1-l, \, \nu \rightarrow \nu,
\eeq
which means that the sets $$D_{++}= D_{+-}, \, D_{-+}= D_{--}$$
provided we allow $l$ to be any integer.

The second duality is more interesting.
Let us note that in the $D_{+-}$ case, when $m=M$, the resulting potential is
\be
\label{dual}
V_M = V^*= x^6 - (\nu-6M)x^2 + \frac{(l+M)(l+M+1)}{x^2}
\eeq
which corresponds to the change
\be
\label{dual2}
\nu \rightarrow \nu^*= \nu - 6M,\quad
l \rightarrow l^* = l+M.
\eeq
We can formally add
\be
 M \rightarrow 
M^* = -M
\eeq
to make it an involution, which is nothing else but the "third spectral equivalence" from \cite{DDT}.

Note that we can allow $M$ to be a negative integer, provided we replace $M$ by $|M|$ in the formula for the degree of the polynomials $P_I$
$$2d = m(|M|-m).$$ This duality means that one can consider the potential (\ref{D+-})
as a result of Darboux transformations applied to both 
$$
V= x^6 - \nu x^2 + \frac{l(l+1)}{x^2} 
$$
and its dual (\ref{dual}). In the second case one should replace the subset $I \subset \{1,2, ..., |M|\}$ by its complement $I^* = \{1,2, ..., |M|\} \setminus I$ (and thus $m$ by $|M|-m$).

\begin{prop}
The results of all possible Darboux transformations applied to the operator (\ref{Sch7})
form one of the 4 sets $D_{\pm \pm}$ described above.
These sets consist of $2^{|M|}$ monodromy-free potentials each and are related by the dualities (\ref{dual1}),(\ref{dual2}). The corresponding Schr\"odinger equations have $|M|$ quasi-rational solutions of the form
$$\psi_i(x) = R_i(x) \exp{(\pm x^4/4)}$$ with some rational functions $R_i(x).$ 
 \end{prop}
 
\medskip

{\bf Example.} We saw above that the operator
$$
L= -D^2 +  x^6 - 7 x^2
$$
has two quasi-polynomial eigenfunctions
$$
\psi_{1,2}= (\pm \sqrt{2} x^2+1) \exp(-x^4/4),
$$
corresponding to 
$\lambda_1 = - 2 \sqrt{2}$ and $\lambda_2 = 2 \sqrt{2}$ respectively. 

The \DT corresponding to $\psi_1$ 
gives a \Schr operator with potential
$$
V_1 = x^6 - 7 x^2 -2 D^2 \ln \psi_1\\
= x^6 -  x^2 + \frac{8 x^2 - 4 \sqrt{2}}{(\sqrt{2} x^2 +1)^2}.
$$
Note that because $\psi_1$ is the ground state of the operator $L$, 
the result is non-singular on the real line. 
The corresponding \Sch equation 
$$
(-D^2 + x^6 -  x^2 + \frac{8 x^2 - 4 \sqrt{2}}{(\sqrt{2} x^2 +1)^2})\phi = \lambda \phi
$$
has two quasi-rational solutions:
$$
\phi_1 = \psi_1^{-1} = \frac{1}{\sqrt{2} x^2 +1}\exp(x^4/4)
$$
with $\lambda = \lambda_1 = - 2 \sqrt{2}$ and
$$
\phi_2 = (D- D \ln \psi_1) \psi_2 = \frac{4\sqrt{2} x}{\sqrt{2} x^2 +1}\exp(-x^4/4)
$$
with $\lambda = \lambda_1 = - 2 \sqrt{2}.$
The first function grows at infinity, and thus has no real spectral 
interpretation, while the second one is a genuine eigenfunction. 
Because it has a zero at $x=0$ it is the first eigenfunction above the 
ground state of the transformed potential. 
Note that the new ground state therefore is not quasi-rational.

Similarly the Darboux transformation corresponding to $\psi_2$ 
gives the potential
$$
V_2 = x^6 - 7 x^2 -2\frac{{\rm d}^2}{{\rm d}x^2}\ln \psi_2 \\
= x^6 -  x^2 +\frac{8 x^2 + 4 \sqrt{2}}{(\sqrt{2} x^2 -1)^2}.
$$

Applying two successive Darboux transformations (which commute) to both $\psi_1$ and $\psi_2$, we come to the potential
$$
V_{1,2} = x^6 - 7 x^2- 2 \frac{{\rm d}^2}{{\rm d}x^2}\ln({\rm
W}(\psi_1,\psi_2))\\
= x^6 + 5 x^2 -2 \frac{{\rm d}^2}{{\rm d}x^2}\ln({\rm
W}(\sqrt{2} x^2+1,\sqrt{2} x^2-1))
$$
$$
= x^6 + 5 x^2 -2 \frac{{\rm d}^2}{{\rm d}x^2}\ln(4 \sqrt{2}x)\\
= x^6 + 5 x^2 +\frac{2}{x^2},
$$
which is of course dual to the initial potential $V=x^6 - 7 x^2.$

More examples led us to the following
\medskip

 {\bf Conjecture.} {\it The polynomials $P_I(x)$ always have simple zeroes, so all the corresponding potentials
$V_I$ have the form (\ref{gen}).}

\medskip

We should note that there are monodromy-free potentials which are 
not of this form, as the following example shows.

Consider a potential of the form
\be
V(x) = x^6 -\nu x^2 + \frac{6}{(x-a)^2} + \frac{6}{(x+a)^2}.
\label{exam6}
\eeq
The trivial monodromy conditions give two relations
$$
6a^5 - 2\nu a - \frac{3}{2a^3}=0, \quad
10a^3 - \frac{3}{8a^5}=0.
$$ 
The solution exists only when
$$
\nu = \pm \frac{51}{4\sqrt{15}}
$$
in which case 
$$a^8=\frac{3}{80}.$$

Note that the corresponding $\nu$ is irrational; therefore such a 
potential can not be a result of Darboux transformations discussed above,
It can not even have quasi-rational solutions, as will follow from the 
results of the next section.

\section{Quasi-rational solutions and Stieltjes relations}

In this section we are going to discuss the relation between two properties:
trivial monodromy and the existence of quasi-rational eigenfunctions. 

We start with a more general situation. Let $V$ be a rational potential of the form
\be
V(x) = P(x) +  \sum_{i=1}^{N}\frac{2}{(x-x_i)^2},
\label{gfor}
\eeq
where $P(x)$ is a polynomial and $x_i$ are distinct.
Assume that the corresponding \Sch equation
$$(-D^2 + V(x))\psi = \lambda \psi$$
has a quasi-rational solution of the form
\be
\label{form}
\psi(x) = \frac{\prod_{j=1}^K (x-y_j)}{\prod_{i=1}^N (x-x_i)} \exp{Q(x)},
\eeq
where $Q(x)$ is another polynomial.
The corresponding logarithmic derivative
$$
f(x) = D \ln \psi(x) = \sum _{j=1}^K \frac{1}{x-y_j} 
- \sum _{i=1}^N \frac{1}{x-x_i} + Q'(x)
$$
must satisfy the Riccati relation
$$
f' + f^2 = V-\lambda.
$$
In particular, the residues at all the poles in the left hand side 
must be zero.
This gives us the following relations between poles $x_i$ 
and zeroes $y_j$ of any such eigenfunction
\bea
\label{Stiel}
\sum _{j=1}^K \frac{1}{x_k-y_j} - \sum _{i \neq k}^N \frac{1}{x_k-x_i} 
+ Q'(x_k)=0, \, k=1,\dots, N\\
\sum _{j\neq l}^K \frac{1}{y_l-y_j} - \sum _{i=1}^N \frac{1}{y_l-x_i} + Q'(y_l)=
0, \, l=1,\dots, K.
\eea
We will call these the {\bf  Stieltjes relations} because in the 
simplest case, when
$Q$ is quadratic and $K=0$, they reduce to the classical 
Stieltjes relations for the zeroes of Hermite polynomials 
(see more on this in \cite{VesSti}).

It turns out that these relations are strong enough to imply the 
trivial monodromy property for the corresponding
potential $V(x).$

\begin{prop}
If the \Sch operator with potential (\ref{gfor}) has a quasi-rational 
eigenfunction of the form (\ref{form}) then this operator is monodromy-free.
\end{prop}

The proof is elementary (cf. Lemma 2 from \cite{VesSti}). Let
\[
f= -\frac{1}{x-x_i} + c_0 + c_1(x-x_i) + c_2(x-x_i)^2+...
\]
be the Laurent expansion of the logarithmic derivative at a pole $x=x_i.$
The $i$-th Stieltjes relation (\ref{Stiel}) means that $c_0=0.$ We have then
\[ f'+f^2 = (-\frac{1}{x-x_i} + c_1(x-x_i) + c_2(x-x_i)^2+...)' 
+ (-\frac{1}{x-x_i} + c_1(x-x_i) + c_2(x-x_i)^2+...)^2\]
\[ = \frac{2}{(x-x_i)^2} - c_1 + (2c_2-2c_2)(x-x_i)+... 
= \frac{2}{(x-x_i)^2} - c_1 + O(x-x_i)^2.\]
Since $f'+f^2 = V-\lambda$, we see that the trivial monodromy condition $V_1=0$ is satisfied.

\begin{cor}
The Stieltjes relations imply the locus equations for the corresponding potential $V(x)$:
\be
\label{loc}
-\sum _{i \neq k}^N \frac{4}{(x_k-x_i)^3} + P'(x_k)=0, \, k=1,\dots, N.
\eeq
\end{cor}

We will give a direct proof of this in the next section where we 
consider the time-dependent version of the Stieltjes relations.

Note that the corresponding polynomial $P(x)$ is defined here simply as 
the polynomial part of the rational function 
\[Q''(x) + (Q')^2 
+ 2Q' \left(\sum _{j=1}^K \frac{1}{x-y_j} - \sum _{i=1}^N \frac{1}{x-x_i}\right)\]
We can be more explicit in the sextic case when $Q(x)=\pm x^4/4$.

\begin{prop}
Suppose that the  \Sch operator with potential (\ref{gfor}) has a 
quasi-rational eigenfunction of the form
\be
\label{form6}
\psi(x) = \frac{\prod_{j=1}^K (x-y_j)}{\prod_{i=1}^N (x-x_i)} \exp{\epsilon x^4/4},
\eeq
where $\epsilon = \pm 1$, then polynomial $P(x)$ has the form
\be
P(x) = x^6 - \epsilon (2N-2K-3)x^2 +2 \epsilon (\sum_j^K y_j - \sum_i^N x_i)x + const
\eeq
In particular the coefficient of the quadratic term
\[
\nu = \pm (2N-2K-3)
\]
is an odd integer.
\end{prop}

Indeed, in that case,
\[
2Q' \left(\sum _{j=1}^K \frac{1}{x-y_j} - \sum _{i=1}^N \frac{1}{x-x_i}\right)=
2\epsilon x^2 \left(\sum _{j=1}^K (1 + \frac{y_i}{x} + \frac{y_i^2}{x^2}+\dots) - \sum _{i=1}^N (1 + \frac{x_i}{x}+ \frac{x_i^2}{x^2} +\dots)\right)
\]
\[
= 2\epsilon (K-N) x^2 
+ 2\epsilon (\sum_j^K y_j - \sum_i^N x_i)x + 2\epsilon (\sum_j^K y_j^2 - \sum_i^N x_i^2) + O(\frac{1}{x}).
\]

Note the existence of a {\bf linear term} 
\[2\epsilon (\sum_j^K y_j - \sum_i^N x_i)x,\] 
which does appear already in the simplest case $K=N=1$, 
as the following example shows.

{\bf Example.} Consider the quasi-rational eigenfunctions of the form
\be
\psi(x) = \frac{x-y_1}{x-x_1} \exp{-x^4/4}
\eeq
The Stieltjes relations 
$$\frac{1}{x_1-y_1}-x_1^3 = 0, \quad -\frac{1}{y_1-x_1}-y_1^3 = 0$$
imply that $x_1^3 = y_1^3,$ so $y_1=\omega x_1$, where $$\omega=\frac{-1\pm \sqrt{3}}{2}$$
is a root of the equation $\omega^2 + \omega + 1 = 0.$
The pole $x_1$ satisfies the equation
$$x_1^4 = \frac{1}{1-\omega}$$
and the corresponding potential is
\be
\label{poten}
V(x) = x^6 - 3x^2 -2(\omega -1)x_1 x.
\eeq

The linear term of course disappears in the symmetric case, for then 
\[\sum_j^M y_j =  \sum_i^N x_i=0.\]
We are going to restrict ourselves to the symmetric case from now on, 
allowing the additional $\frac{l(l+1)}{x^2}$ term at zero:
\be
\label{ge}
V(x) = x^6 - \nu x^2 + \frac{l(l+1)}{x^2} + \sum_{i=1}^{N}\frac{2}{(x-x_i)^2}.
\end{equation}
The existence of quasi-rational solutions of the form
\be
\label{formpm}
\psi(x) = x^{\mu}\frac{\prod_{j=1}^K (x-y_j)}{\prod_{i=1}^N (x-x_i)} \exp{(\pm x^4/4)},
\eeq
with $\mu = -l$ or $\mu = l+1$, leads to the following Stieltjes relations:
\bea
\label{Stiel2}
\sum _{j=1}^K \frac{1}{x_k-y_j} - \sum _{i \neq k}^N \frac{1}{x_k-x_i}  \pm x_k^3 + \frac{\mu}{x_k} =0, \, k=1,\dots, N\\
\sum _{j\neq l}^K \frac{1}{y_l-y_j} - \sum _{i=1}^N \frac{1}{y_l-x_i} \pm y_l^3 +\frac{\mu}{y_l} =0, \, l=1,\dots, M.
\eea
Exactly as before, they imply the locus relations
\begin{equation}
\label{locus}
 \sum_{j\ne i}^{N}\frac{2}{(x_i - x_j)^3} + \frac{l(l+1)}{x_i^3} 
 + \nu x_i - 3x_i^5 =0, \quad i=1,\dots, N,
\end{equation}
with 
\[\nu = \pm(2N-2K-2\mu-3).\]
Note that this is true for all values of $l$ (and thus $\mu$), 
not necessarily integers.

The question is whether all solutions of the locus system can be found 
in such a way.
In other words: is it true that any even monodromy-free potential of 
the form (\ref{ge}) must have a quasi-rational eigenfunction?

The answer is `No' as the simplest example of the potential
$$V(x) = x^6 - (2l+1)x^2 + \frac{l(l+1)}{x^2}$$
shows (see section 3 above). A more interesting example is given by
$$U = x^6  - x^2  + 2/(x-a)^2 + 2/(x+a)^2.$$
The locus relation is
$$12 a^8 - 4 a^4 - 1= 0,$$
which gives us two possibilities: either $a^4 = 1/2$ or $a^4 = -1/6.$
Since $\nu = \pm(4-2K-2\mu -3) = \pm(1-2K-2\mu)=1$
and $\mu=0$ or $\mu = 1$ one can see that $K$ must be either zero or 1 and we have two possible quasi-rational eigenfunctions:
$$\psi=\frac{x-b}{x^2-a^2} \exp{(-x^4/4)}$$
and
$$\psi = \frac{1}{x^2-a^2} \exp{(x^4/4)}.$$
In the first case the Stieltjes relations are
$$
\frac{1}{a-b} - \frac{1}{2a} - a^3=0, \quad \frac{1}{-a-b} + \frac{1}{2a} + a^3=0,
$$
implying $b=0$ and $a^4=1/2.$
In the second case we have just $$- \frac{1}{2a} + a^3=0,$$
which give the same $a^4=1/2.$
Thus we see that the solution $a^4 = -1/6$ of the locus conditions is inconsistent
with the Stieltjes relations, and thus the corresponding monodromy-free potential
has no quasi-rational eigenfunctions.

We note that the potentials with $a^4=1/2$ are exactly those given by 
the Darboux transformations applied to 
$V=x^6-7x^2$ (see example in the section 3 above). 
This raised the following natural question.
Suppose that the monodromy-free potential (\ref{ge}) has a quasi-rational
function of the form (\ref{formpm}). Is it true that it is a result of 
Darboux transformation applied to the operators (\ref{Sch7}) ?

We will show that this answer is negative as well. This will follow from
the results of the next section, where we investigate the
corresponding Darboux set in more detail.

\section{Geometry and arithmetic of the Darboux set}

By the {\it Darboux set} $\mathbf D$, we will mean
all possible results of Darboux transformations applied to
\be
\label{new}
V=x^6 - \nu x^2 + \frac{l(l+1)}{x^2}.
\eeq
It is the union of the sets $D_{\pm\pm}$ described in section 3. 
We will allow $l_0$ (and the corresponding $\mu$ in (\ref{form})) to be 
arbitrary (not necessarily integer) parameters in this section.
The corresponding Schr\"odinger operators will have some monodromy at $x=0$
in that case but will be monodromy-free elsewhere. This does not affect the
relation between the locus and Stieltjes algebraic systems shown in the 
previous section.

We know that the the operator with the potential (\ref{new}) has a 
quasi-rational solution (\ref{form})
only if some of the numbers $$M = \frac{\pm \nu \pm (2l+1)}{4}$$ are integers;
in that case, as shown in section 3, 
the number of such solutions is equal to $|M|$.

It is convenient to represent these cases on the $(l, \nu)$-plane as the grid $\Gamma$  of  the lines
$$\nu \pm(2l+1) =4M_{\pm}$$ with integer $M.$ 

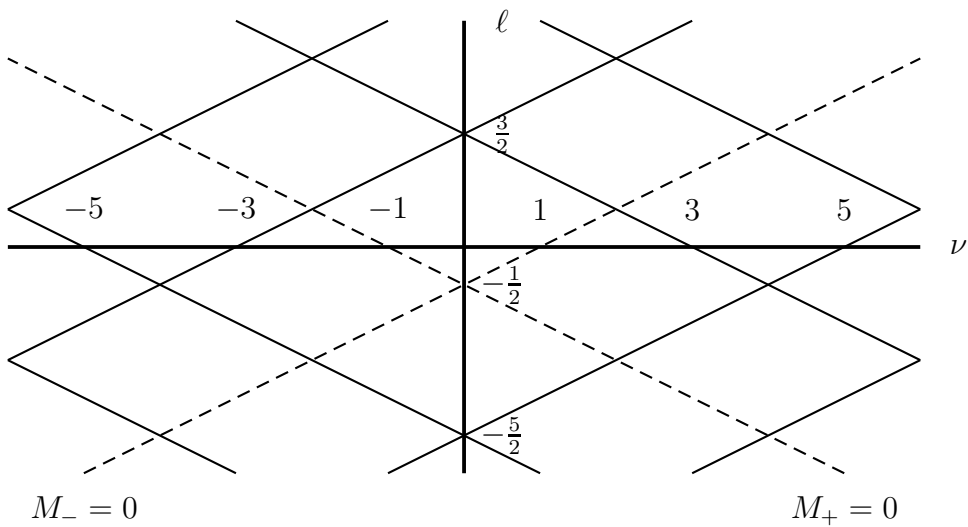
\bigskip
\bigskip
\bigskip

\begin{center}
 \begin{figure}[ht] 
 \begin{pspicture}(0,-3.5)(12,3.5) 
 \psline[linewidth=0.5mm](0,0)(12,0)
 \psline[linewidth=0.5mm](6,-3)(6,3)
 \rput(11,-3.5){$M_{+}=0$}
 \psline[linestyle=dashed](11,-3)(0,2.5)
 \psline[linestyle=dashed](1,-3)(12,2.5)
 \psline(3,3)(12,-1.5)
 \psline(9,3)(0,-1.5)
 \psline(0,0.5)(7,-3)
 \psline(12,0.5)(5,-3)
 \psline(0,0.5)(5,3)
 \psline(12,0.5)(7,3)
 \psline(0,-1.5)(3,-3)
 \psline(12,-1.5)(9,-3)
 \rput(1,-3.5){$M_{-}=0$}
 \rput(12.5,0){$\nu$}
 \rput(6.5,3){$\ell$}
 \rput(1,0.5){$-5$}
 \rput(3,0.5){$-3$}
 \rput(5,0.5){$-1$}
 \rput(7,0.5){$1$}
 \rput(9,0.5){$3$}
 \rput(11,0.5){$5$}
 \rput(6.5,-0.5){$-\frac{1}{2}$}
 \rput(6.5,-2.5){$-\frac{5}{2}$}
 \rput(6.5,1.5){$\frac{3}{2}$}
 \end{pspicture} 
 \end{figure}
\end{center}

\bigskip
\bigskip
\bigskip

Strictly speaking, when $M_{\pm}=0$ we do not have quasi-rational 
solutions but we want anyway
to include the corresponding two lines (drawn dashed) 
$$
\nu \pm (2l+1) =0,
$$
which we call {\it special}. 
The number of quasi-rational solutions is proportional to the distance 
from these lines. 

Note also that all the intersection points of the grid have $l$ 
coordinates which are {\it half-integers}. In these cases we can have 
all four different types of quasi-rational solutions simultaneously, 
while at all other points in the grid we have only two of them.

Now let us consider the Darboux descendants of these potentials. 
They have the form
$$
V_I=x^6 - \nu x^2 + \frac{l(l+1)}{x^2} - 2 D^2 \ln P_{2d}(x)
$$
with the parameters $\nu$ and $l$ different from the original form (\ref{new}).

Our first claim that these new parameters lie on the same grid $\Gamma$.
One can easily check this in all the cases $D_{\pm \pm}.$
For example the Darboux transformations in $D_{+-}$ case, change
$$\nu \rightarrow \nu -6m, \quad l \rightarrow l+m,$$
so $$M = \frac{\nu -(2l+1)}{4} \rightarrow \frac{\nu -(2l+1)}{4} - 2m = M-2m,$$
where $m$ is an integer between 1 and $|M|.$
So we see that $M$ remain integers, but with modulus $|M|$ decreasing,
we moved on the grid closer to the special lines.
One can check that the same change
\be
\label{change}
M \rightarrow M-2m
\eeq
is true for all other types $D_{\pm\pm}$ as well.

Let us look now at the degree of the corresponding polynomials 
$P_{2d}(x)$, which is the same as the number $N$ of the poles in the 
corresponding potential. We know that in all the cases it has the 
product form
\be
\label{prod}
N=2d=2m(|M_0|-m),
\eeq
where 
\[M_0=\frac{\pm \nu_0 \pm (2l_0+1)}{4}\] 
corresponds to the initial operator
\be
\label{new0}
V=x^6 - \nu_0 x^2 + \frac{l_0(l_0+1)}{x^2}.
\eeq

This means that the arithmetic of $N$ plays a substantial role here.
In particular if the number of poles is $2p$ where $p$ is prime then 
this implies that
corresponding $m$ must be either 1 or $p=|M_0|-1$ 
(which is actually a dual case to $m=1$, see section 3). 
So we can claim that such a potential must be a result of just one 
Darboux transformation and
the corresponding parameters $(l,\nu)$ must lie on the lines
\be
\label{plines}
\nu \pm (2l+1) = 4(p-1).
\eeq
At a generic point on these lines the initial parameters $\nu_0, \l_0$ 
are determined uniquely, so we have exactly $p+1$ such potentials 
depending on the choice of one of the corresponding $|M_0|=p+1$ 
quasi-polynomial eigenfunctions.

If $d$ is not prime then the number of Darboux steps $m$ could be any 
divisor of $d = mk,$ where $k$ can be assumed larger than $m.$
In that case $|M_0|=k+m$ and we have
\[\frac{(m+k)!}{m! k!}\] 
potentials on the lines
\be
\label{plines}
\nu \pm (2l+1) = 4(k-m).
\eeq

Let us come back now to the question whether any monodromy-free 
operator with a quasi-rational eigenfunction belongs to the Darboux set 
or not. The following simple example shows that the answer in general 
is negative.

{\bf Example.} We consider the potentials of the form
\[U = x^6  - 3x^2  + 2/(x-a)^2 + 2/(x+a)^2\]
and check whether they have quasi-rational functions of the form
\[
\psi(x)=\frac{x^2-b^2}{x^2-a^2}\exp{(-x^4/4)}
\]
with $a$ and $b$ different from zero.
The Stieltjes relations here,
\[
\frac{1}{2b} - \frac{2b}{b^2-a^2} - b^3=0, \quad -\frac{1}{2a} + \frac{2a}{a^2-b^2} - a^3=0,
\]
are equivalent to
\[
a^4 - b^4 = 1, \quad a^4 + 4 a^2 b^2 + b^4 =0.
\]
This gives 8 solutions 
\[a^4 = \frac{1}{1-\epsilon^2},\quad 
b^2 = \epsilon a^2,\] 
where $\epsilon$ satisfies the quadratic equation
$\epsilon^2 + 4 \epsilon + 1 =0$, giving:
\[\epsilon = -2 \pm \sqrt{3}.\]

Now we claim that the corresponding potentials cannot be the result of 
Darboux transformations.
Indeed in that case $d=1$, so $m$ must be 1 and $|M_0|=2$. On the other 
hand, since
$l=0, \nu =3$ we have $M=\frac{3+1}{4}=1$, which must be equal to 
$M_0-2m = M_0-2,$
which is clearly impossible.

\section{Time-dependent Stieltjes relations and special solutions of 
the Calogero-Moser problem}

We now consider the {\em time-dependent} \Schr equation 
\begin{equation}
i \frac{\dd}{\dd t}\psi=\half(  - \frac{\partial^2}{\partial x^2}+V(x))
\psi. \label{TDSch}
\end{equation}
If $V(x,t)=  x^6 - \nu x^2 +\frac{l(l+1)}{x^2}$, then this has a linear space
of dimension $M$ 
of quasi-rational solutions, with exponential time-dependence: if $\psi^{i}$ satisfies
$$
 (-  \frac{\dd^2}{\dd x^2}+V(x))\psi^i = \l_i \psi ^i,
$$
then $\psi_{\bf c} = \sum_{j=1}^M c_j \psi ^j \exp(-i \l_i t/2)$ evidently
solves (\ref{TDSch}). This solution, which has simple but non-trivial time-dependence,
may be used, as above, to construct a \DT
$$
V_{\bf c_1,\ldots,c_M} = V
- 2\frac{\dd^2}{\dd x^2}\ln({\rm W}_{j=1}^M( \psi^{(0)}_{\bf c_j}).
$$
The moving poles of the new potential must thus satisfy the monodromy conditions;
in particular they must solve the corresponding \CM system.

As an example let us consider the  $N=2$ case.
Putting exponential time-dependence into the eigenfunctions discussed above,
we construct a $2$-dimensional space of \qpoly solutions of the time-dependent
\Schr equation:
$$
i \frac{\dd}{\dd t}\psi= (-\frac{\dd^2}{\dd x^2}+ x^6 -7x^2)
\psi,
$$
namely
$$
\psi_{\bf c} = (c_1 (\sqrt{2} x^2+1){\rm e}^{i\sqrt{2}t}
+c_2 (\sqrt{2} x^2-1){\rm e}^{-i\sqrt{2}t})\exp(-x^4/4).
$$
The \DT associated with this leads to a new potential
\begin{eqnarray*}
V^{(1)}_{\bf c} =  x^6 - x^2-2\frac{\dd^2}{\dd x^2}\ln(\psi_{\bf c})
\\
= x^6 -x^2 + 4\frac{x^2 + X^2(t)}{(x^2 - X^2(t))^2},
\end{eqnarray*}
where
$$
X^2(t) 
= -\frac{c_1 {\rm e}^{i\sqrt{2}t} -c_2 {\rm e}^{-i\sqrt{2}t}}
{\sqrt{2}(c_1 {\rm e}^{i\sqrt{2}t} +c_2 {\rm e}^{-i\sqrt{2}t})}
$$
$$= - i \frac{\tan(\sqrt{2}(t-t_0))}{\sqrt{2}}.$$
We may easily check in this simple case that the points 
$$x_1(t) =X(t),$$
$$x_2(t) = - X(t)$$ 
indeed satisfy the \CM system
$$
 \ddot{x}_i= \sum_{j=1,j\ne i} \frac{2}{(x_i-x_j)^3} - 3 x_i^5 + x_i.
$$
From the locus condition, such a result in fact holds for all 
potentials of this kind,
as will be shown more directly below, 
where for simplicity we assume the parameter $l=0.$

As in the time-independent case, \DTs take a \qpoly solution
of the time-dependent \Schr equation with an even sextic potential
$$
V^{(0)}(x) =  x^6 - (2n+3) x^2
$$
into a \qrat solution 
$$
\psi^{(m)}_{\bf c_1,\ldots, c_m;c_{m+1}} 
= \frac{{\rm W}_{j=1}^{m+1}( \psi^{(0)}_{\bf c_j})}{{\rm W}_{j=1}^{m}(\psi^{(0)}_{\bf c_j})},
$$
which satisfies the \Schr equation with the rational potential
\begin{eqnarray*}
V^{(m)}_{\bf c_1,\ldots,c_m} =  x^6 - (2n+3) x^2
- 2\frac{\dd^2}{\dd x^2}\ln({\rm W}_{j=1}^m( \psi^{(0)}_{\bf c_j})\\
= x^6 - (2n+3-6m) x^2
- 2 \frac{\dd^2}{\dd x^2}\ln({\rm W}_{j=1}^m( \phi^{(0)}_{\bf c_j}).
\end{eqnarray*}
Here, as before, the $\phi^{(0)}_{\bf c_j}$ are the rational factors of the
corresponding \qrat solutions $\psi^{(0)}_{\bf c_j}$.

As before, this solution may be described in terms of its poles and 
zeroes as 
$$
\psi^{(m)}=\frac{ \prod_{j=1}^Z (x-x_j)^{\a_j}}{ \prod_{k=1}^P (x-\tilde{x}_k)^{\beta_k}}\exp(-x^4/4)\exp(i
f(t)),
$$
where the multiplicities $\a_j$ and $\beta_k$ are all positive integers.
Here the points $x_k$, $\tilde{x}_k$ are functions of time, while the time-dependent
phase $f(t)$ is not yet determined.  
We identify the
numerator and denominator of the rational part with the polynomial 
factors of the two Wronskians.
$$
W_{m+1}=\prod_{j=1}^Z (x-x_j)^{\a_j} = 
{\rm W}_{j=1}^{m+1}( \phi^{(0)}_{\bf c_j}),
$$
$$W_m=\prod_{k=1}^P (x-\tilde{x}_k)^{\beta_k} = 
{\rm W}_{j=1}^{m}(\phi^{(0)}_{\bf c_j}).
$$
We note that the numerator and denominator may have common factors. The 
\Sch equation
is now
$$ i\frac{\dd }{\dd t}\frac{W_{m+1}}{W_m}- \dot{f} \frac{W_{m+1}}{W_m}= $$
$$ \left( -\half(\frac{\dd}{\dd x}-x^3)^2 
+\half x^6 -(N-3 m+3/2)x^2
-\frac{\dd^2}{\dd x^2} \ln({W_m})\right) \frac{W_{m+1}}{W_m},$$
and the Hirota form is hence:
$$
i \frac{\dd W_{m+1} }{\dd t}W_m-iW_{m+1}\frac{\dd W_{m} }{\dd t}-\dot{f}W_{m+1}W_m
$$
$$+\half 
(\frac{\dd^2 W_{m+1}}{\dd x^2}W_m-2\frac{\dd W_{m+1}}{\dd x}\frac{\dd W_m}{\dd x}
+W_{m+1}\frac{\dd^2 W_{m}}{\dd x^2})$$
$$- x^3 (\frac{\dd W_{m+1}}{\dd x}W_m-W_{m+1}\frac{\dd W_{m}}{\dd x})$$
$$=x^2 (N-3m) W_{m+1}W_m.
$$
Now, on dividing the whole equation by $W_m W_{m+1}$, this becomes:
$$
-i \sum_{j=1}^Z \frac{\a_j \dot{x}_j}{x-x_j} 
+i \sum_{k=1}^P \frac{\beta_k \dot{\tilde{x}}_k}{x-\tilde{x}_k} - \dot{f}$$
$$+\half \left(\sum_{j=1}^Z \sum_{k=1,k\ne j}^Z \frac{\a_j \a_k}{(x-x_j)(x-x_k)}
-2\sum_{j=1}^Z \sum_{k=1}^P \frac{\a_j \beta_k}{(x-x_j)(x-\tilde{x}_k)}
+ \sum_{j=1}^P \sum_{k=1,k\ne j}^P \frac{\beta_j \beta_k}{(x-\tilde{x}_j)(x-\tilde{x}_k)}\right)$$
$$+x^3(\sum_{j=1}^Z \frac{\a_j }{x-x_j} 
- \sum_{k=1}^P \frac{\beta_k}{x-\tilde{x}_k}) = x^2 (N-3m).$$
Matching coefficients of $x^2$, we again get $\sum_{j=1}^Z \a_j-\sum_{j=1}^P \beta_j=N-3m$.
The coefficients of $x$ must all vanish, by parity, and the coefficient
of $x^0$ can now be balanced by requiring
$$ -\dot{f} + \sum_{j=1}^Z \a_j x_j^2
-\sum_{j=1}^P \beta_j \tilde{x}_k^2=0.
$$
It remains to check the residues
at the zeroes $x_j$ and poles $\tilde{x}_k$ which give, respectively:
\be
-i \a_j \dot{x}_j + \sum_{k=1,k\ne j}^Z \frac{\a_j \a_k}{(x_j-x_k)}- \sum_{k=1}^P \frac{\a_j \beta_k}{(x_j-\tilde{x}_k)}+
x_j^3 \a_j =0,\label{SDzeroes}
\eeq
and
\be
i \beta_k \dot{\tilde{x}}_k - \sum_{j=1,j\ne k}^Z \frac{\a_j \beta_k}{(\tilde{x}_k-x_j)}+ \sum_{j=1,j\ne
k}^P \frac{\beta_j \beta_k}{(\tilde{x}_k-\tilde{x}_j)}+
\tilde{x}_k^3 \beta_k =0.\label{SDpoles}
\eeq
These are the dynamical Stieltjes relations we seek.

To simplify the notation, if we look at the union of the sets of poles and zeroes, points $z_i$,
each with a positive or negative exponent $\gamma_i$,
$$ z_i=x_i,\quad \gamma_i=\alpha_i, \quad i=1,\ldots,Z,$$
$$ z_i=\tilde{x}_{i-Z},\quad \gamma_i=-\beta_{i-Z}, \quad i=Z+1,\ldots,Z+P,
$$
then this set of relations simplifies:
\be
 \dot{z}_j =-i \sum_{k=1,k\ne j}^{Z+P} \frac{\gamma_k}{(z_j-z_k)}-i z_j^3,\label{Sdyn}
\eeq
It is instructive to differentiate these relations with respect to time.
We find:
$$
 \ddot{z}_j =i \sum_{k=1,k\ne j}^{Z+P} \frac{\gamma_k}{(z_j-z_k)^2}(\dot{z}_j-\dot{z}_k)
 -3i z_j^2  \dot{z}_j.
$$
On substituting (\ref{Sdyn}) into this, we obtain
\begin{eqnarray*}
 \ddot{z}_j =i \sum_{k=1,k\ne j}^{Z+P} \frac{\gamma_k}{(z_j-z_k)^2}\\
 (-i \sum_{l=1,l\ne j}^{Z+P} \frac{\gamma_l}{(z_j-z_l)}-i z_j^3
  +i \sum_{l=1,l\ne k}^{Z+P} \frac{\gamma_l}{(z_k-z_l)}+i z_k^3 )\\
 -3i z_j^2(-i \sum_{l=1,l\ne j}^{Z+P} \frac{\gamma_k}{(z_j-z_l)}-i z_j^3).
\end{eqnarray*}
This may be rearranged as
$$
\ddot{z}_j =\sum_{k=1,k\ne j}^{Z+P}\frac{\gamma_k(\gamma_k+\gamma_j)}{(z_j-z_k)^3}$$
$$ +\sum_{k=1,k\ne j}^{Z+P}\sum_{l=1,l\ne j,k}^{Z+P}
 \frac{\gamma_k}{(z_j-z_k)^2}(
 \frac{\gamma_l}{(z_j-z_l)}+ z_j^3-\frac{\gamma_l}{(z_k-z_l)}- z_k^3)$$
$$  -3 z_j^2(\sum_{l=1,l\ne j}^{Z+P} \frac{\gamma_k}{(z_j-z_l)}-i z_j^3).$$
This further simplifies, after some elementary manipulation, to
$$
\ddot{z}_j =
 \sum_{k=1,k\ne j}^{Z+P}\frac{\gamma_k(\gamma_k+\gamma_j)}{(z_j-z_k)^3}-3
z_j^5\\
-\sum_{k=1,k\ne j}^{Z+P}\sum_{l=1,l\ne j,k}^{Z+P}
  \frac{\gamma_k\gamma_l}{(z_j-z_k)(z_j-z_l)(z_k-z_l)}
  -\sum_{k=1,k\ne j}^{Z+P} \gamma_k(2z_j+ z_k).$$
Now in this long expression the double sum 
$$\sum_{k=1,k\ne j}^{Z+P}\sum_{l=1,l\ne j,k}^{Z+P}
  \frac{\gamma_k\gamma_l}{(z_j-z_k)(z_j-z_l)(z_k-z_l)}$$ 
  is identically zero.
  The final term 
  $$\sum_{k=1,k\ne j}^{Z+P} \gamma_k(2z_j+ z_k)$$ 
is equal
to 
$$z_j(2\sum_{k=1}^{Z+P} \gamma_k - 3\gamma_j)+\sum_{k=1}^{Z+P} \gamma_k z_k).$$
In the special case where all zeroes and poles are simple, 
$\gamma_k=\pm 1$, and
$\sum_{k=1}^{Z+P} \gamma_k=Z-P$, so this term becomes
$z_j (2 (Z-P) -3 \gamma_j)$. Further,
from the quadratic term above we saw $Z-P = N-3m$. The term $\sum_{k=1}^{Z+P} \gamma_k z_k$
is supposed to be identically zero by parity. 

Thus, remarkably, the system decouples, giving:
\be
\ddot{x}_j =\sum_{k=1,k\ne j}^{Z}\frac{2}{(x_j-x_k)^3}-3
x_j^5
-x_j (2 (Z-P) -3).
\eeq  
for the zeroes, and
\be
\ddot{\tilde{x}}_j =\sum_{k=1+Z,k\ne j}^{Z+P}\frac{2}{(\tilde{x}_j-\tilde{x}_k)^3}-3
\tilde{x}_j^5
-\tilde{x}_j (2 (Z-P) +3).
\eeq
for the poles.

\begin{prop}
The Stieltjes relations (\ref{SDzeroes}, \ref{SDpoles}) with $\alpha_i = \beta_j=1$ imply
that the zeroes and poles of solutions of the non-stationary \Sch equation satisfy the uncoupled  \CM systems 
in different sextic potentials with the Hamiltonians
$$
H = \half \sum_{i=1}^Z p_i^2 +  \half \sum_{i=1}^Z x_i^6 -  \half \sum_{i=1}^Z
(2 (Z-P)-3) x_i^2 +\half \sum_{i=1}^Z \sum_{j=1,j\ne i}^Z  \frac{1}{(x_i-x_j)^2}
$$
for the zeroes, and
$$
\tilde{H} = \half \sum_{i=1}^P \tilde{p}_i^2 +  \half \sum_{i=1}^P \tilde{x}_i^6 
-  \half \sum_{i=1}^P (2 (Z-P)+3) \tilde{x}_i^2 
+\half \sum_{i=1}^P \sum_{j=1,j\ne i}^P  \frac{1}{(\tilde{x}_i-\tilde{x}_j)^2}
$$
for the poles.
\end{prop}

Thus the first order equations (\ref{Sdyn}) again imply that the simple
poles and zeroes satisfy separate \CM systems, in even sextic potentials,
but with different $O(x^2)$ terms. We may thus regard (\ref{Sdyn})
as being a \BT for \CM systems of this type. This does not imply the system
is integrable however, for the \BT does not depend on an arbitrary constant.
Further, it can only be iterated finitely many times.
 
Alternatively, the system of dynamical Stieltjes relations
may be understood as a canonical transformation. Taking all the poles and
zeroes to be simple, if the  momentum variable conjugate to the zero $x_j$
is $p_j$, and that conjugate to $\tilde{x}_j$ is $\tilde{p}_j$,
the relations read:
\be
p_j =-i \sum_{k=1,k\ne j}^Z \frac{1}{(x_j-x_k)}
+i\sum_{k=1}^P \frac{ 1}{(x_j-\tilde{x}_k)}
-i x_j^3  ,\label{Czeroes}
\eeq
and
\be
 \tilde{p}_k =  i \sum_{j=1}^Z \frac{1 }{(\tilde{x}_k-x_j)}-i \sum_{j=1,j\ne
k}^P \frac{1 }{(\tilde{x}_k-\tilde{x}_j)}-i \tilde{x}_k^3  =0.\label{Cpoles}
\eeq
If we define
\bea
S = -i \log\left(\frac{
 \prod_{j=1}^Z\prod_{k=1,k>j}^Z (x_j-x_k)\prod_{m=1}^P\prod_{n=1,n> m}^P (\tilde{x}_m-\tilde{x}_n)}
{\prod_{j=1}^Z\prod_{k=1,k\ne j}^P (x_j-\tilde{x}_k)}\right) \nonumber\\
-i\sum_{j=1}^Z\frac{x_j^4}{4}+i\sum_{k=1}^P\frac{\tilde{x}_j^4}{4},
\eea
then the Stieltjes relations (\ref{Czeroes}, \ref{Cpoles}) become:
\bea
p_j =\pdd{S}{x_j},\\
\tilde{p_j} =-\pdd{S}{\tilde{x}_j}.
\eea
They are thus seen to give a canonical transformation with 
generating function $S$. Note that dropping  $i$ in these formulae 
corresponds to a change of sign in the potential and thus to an 
attractive, rather than repulsive, pairwise potential.

This transformation is straightforward when the numbers 
$Z$ and $P$ of  zeroes and poles are equal, but otherwise we may 
still understand it as generating a mapping
from simple solutions of a \CM system, with few degrees of freedom, 
to more complicated solutions, with more degrees of freedom - 
of course such solutions will be special, with some
algebraic relations between the new coordinates and momenta. The rational
potentials constructed above correspond to the most special ones of all 
these, which are connected
by a finite chain of such canonical transformations to the `empty' \CM 
system with no particles at all. 
Since it is elementary to show that this transformation
preserves the Hamiltonian, $H=\tilde{H}$, all the solutions of \CM systems
constructed here, related by finitely many such transformations to one
with no degrees of freedom, must have zero energy. 
We believe that the same should be true for other Inozemtsev integrals 
as well. 

One immediate consequence of the fact that the dynamical Stieltjes relations
imply that the simple poles and zeroes separately satisfy \CM equations 
is that the analogous stationary Stieltjes relations must imply that 
the poles and zeroes satisfy stationary \CM equations in the 
appropriate potentials; these are precisely the locus conditions for 
these potentials.

\section{Concluding remarks}

We have seen that the monodromy-free potentials with sextic growth are 
described by the algebraic locus equations and are related to 
equilibria of the corresponding version of the Calogero-Moser problem, 
which is known to be integrable \cite{Inoz}. So, is the locus system 
(\ref{locus}) `integrable' ? In which sense ? 

We have shown that Darboux transformations are not enough to describe 
the whole locus, even if we assume the existence of quasi-rational 
solutions. The geometry and arithmetic of the corresponding potentials 
seems to be quite interesting as well as the algebraic geometry of the 
corresponding (zero-dimensional) affine varieties. In particular, the 
Stieltjes relations can be considered as correspondences (in the sense 
of algebraic geometry) between two different locus varieties, which is 
a partially defined multivalued algebraic map. Although they can be 
considered as natural analogues of the B\"acklund transformations their 
existence is not enough for `integrability' and holds for any growth at infinity
(see section 6 above).

But the general structure of the sextic locus still remain largely unclear.
In relation with Inozemtsev's results \cite{Inoz} we note that
the general solution of the Calogero-Moser system in a sextic potential 
is given in terms of some matrix analogues of the elliptic functions 
and are expressed in terms of the high genus Riemann $\theta$-functions.
This means that an elementary description of all corresponding 
equilibria seems to be unlikely in agreement with our result.

In that sense the situation looks similar to the case of the elliptic 
potentials studied by  Gesztesy and Weikard \cite{GW}, who showed that 
all the corresponding potentials must be finite-gap in the sense of 
\cite{DMN}. The problem of finding an effective description of the 
elliptic finite-gap potentials was first raised by S.P. Novikov in the 
70s, who was inspired by the first examples found in his work with 
Dubrovin \cite{DN}. This area was revitalised after the appearance of 
the famous Trebich-Verdier preprint \cite{TV}, which showed that there 
are examples, which can not be found using the KdV flows from the 
classical Lam\'e potentials $$V(x) = m(m+1)\wp(x),$$ where $\wp(x)$ is 
the Weierstrass elliptic function. 
In spite of the numerous efforts in this direction (see e.g.
Enolskii-Eilbeck \cite{EE}, Taimanov \cite{T}, Smirnov \cite{S}) the 
problem seems to be far from a satisfactory solution.

Our sextic problem can be considered as a simple model for this 
important problem and thus deserves further investigation.
It could also very interesting from the analytic-differential point of 
view, see the recent paper \cite{EGS} by Eremenko, Gabrielov and Shapiro. 

\section{Acknowledgements}

This work has been partially supported by the
European Union through the FP6 Marie Curie RTN ENIGMA (Contract
number MRTN-CT-2004-5652). We have also received support from the European
Science Foundation through the MISGAM programme. 

The work of A.V. was also partially supported by the EPSRC (grant EP/E004008/1).

\end{document}